\documentclass{sig-alternate}

\usepackage[utf8]{inputenc}
\usepackage[T1]{fontenc}
\usepackage{textcomp}
\usepackage[english]{babel}
\usepackage{subcaption}
\usepackage{microtype}
\usepackage{graphicx}
\usepackage{tabulary}
\usepackage{booktabs}
\usepackage{csquotes}
\usepackage[backend=bibtex,natbib=true,style=numeric,maxcitenames=1,maxbibnames=100]{biblatex}
\usepackage{url}
\usepackage{times}
\usepackage{amstext}
\PassOptionsToPackage{fleqn}{amsmath}           
 \usepackage{amsmath}
\usepackage{flushend}
\usepackage{enumitem}
\setlist{nolistsep}

\usepackage[bookmarks,linkcolor=black]{hyperref} 
\hypersetup{
  pdfauthor = {},
  pdftitle = {},
  pdfsubject = {},
  pdfkeywords = {},
  colorlinks=true,
  linkcolor= black,
  citecolor= black,
  pageanchor=true,
  urlcolor = black,
  plainpages = false,
  linktocpage
}
\newfont{\mycrnotice}{ptmr8t at 7pt}
\newfont{\myconfname}{ptmri8t at 7pt}
\let\crnotice\mycrnotice%
\let\confname\myconfname%

\newcommand{\ie}{\emph{i.\,e.}}

\newcommand{\eg}{\emph{e.\,g.}}
 
\newcommand{\wrt}{with respect to }
\usepackage{gensymb}

\usepackage{ifthen}
\newboolean{inthesis}
\setboolean{inthesis}{false}

{\ifthenelse{ \boolean{inthesis} }{}{}
{\ifthenelse{ \boolean{inthesis} }{\newcommand{\bigtable}{sidewaystable}}{\newcommand{\bigtable}{table*}}
{\ifthenelse{ \boolean{inthesis} }{}{}
{\ifthenelse{ \boolean{inthesis} }{}{}
{\ifthenelse{ \boolean{inthesis} }{\newcommand{\spara}[1]{\subsection{#1}}}{\newcommand{\spara}[1]{\smallskip\noindent{\bf #1.}}}

\usepackage{xpatch}

\xpatchbibmacro{name:andothers}{%
  \bibstring{andothers}%
}{%
  \bibstring[\emph]{andothers}%
}{}{}

\addtolength{\dblfloatsep}{-5mm}
\addtolength{\dbltextfloatsep}{-5mm}

\numberofauthors{3}
\author{
\alignauthor Eduardo Graells-Garrido\thanks{Corresponding author: \url{eduardo.graells@telefonica.com}. Work carried out while the first author was a PhD student at Universitat Pompeu Fabra, Barcelona, Spain.} \\
\affaddr{Telefónica I+D} \\
\affaddr{Santiago, Chile} \\
\alignauthor Mounia Lalmas \\
\affaddr{Yahoo Labs} \\
\affaddr{London, UK} \\
\alignauthor Ricardo Baeza-Yates \\
\affaddr{Yahoo Labs} \\
\affaddr{Sunnyvale, USA} \\
}

\title{Finding Intermediary Topics \\ Between People of Opposing Views: A Case Study}

\addbibresource{virtual_portraits.bib}

\makeatletter
\def\@copyrightspace{\relax}
\makeatother

\begin{document}

\maketitle

\begin{abstract}
In micro-blogging platforms, people can connect with others and have conversations on a wide variety of topics. 
However, because of homophily and selective exposure, users tend to connect with like-minded people and only read agreeable information.
Motivated by this scenario, in this paper we study the diversity of intermediary topics, which are latent topics estimated from user generated content. These topics can be used as features in recommender systems aimed at introducing people of diverse political viewpoints.
We conducted a case study on Twitter, considering the debate about a sensitive issue in Chile, where we quantified homophilic behavior in terms of political discussion and then we evaluated the diversity of intermediary topics in terms of political stances of users. 
\end{abstract}


\section{Introduction}
Social research has shown that, while everyone indeed has a voice, people tend to listen and connect only to those of similar beliefs in political and ideological issues, a cognitive bias known as homophily \cite{mcpherson2001birds}.
This bias is present in many situations, and it can be beneficial, as communication with culturally alike people is easier to handle. 
However, the consequences of homophily in ideological issues are prominent, both off- and on-line.
On one hand, groups of like-minded users tend to disconnect from other groups, polarizing group views. 
On the other hand, Web platforms recommend and adapt content based on interaction and network data of users, \ie, who is connected to them and what they have liked before. 
Because algorithms want to maximize user engagement, they recommend content that reinforces the homophily in behavior and display only agreeable information.
Such biased reinforcement, in turn, makes computer systems to recommend even more polarizing content, confining users to \emph{filter bubbles} \cite{pariser2011filter}. 

One way to improve the current situation is to motivate users to read challenging information, or to motivate a change in behavior through recommender systems. 
However, this ``direct'' approach has not been effective as users do not seem to value political diversity or do not feel satisfied with it \cite{an2013individuals}, a result explained by \emph{cognitive dissonance} \cite{festinger1962theory}, a state of discomfort that affects persons confronted with conflicting ideas, beliefs, values or emotional reactions. 
Conversely, \citet{graells2014people} proposed an ``indirect'' approach, by taking advantage of partial homophily to suggest similar people, where similarity is estimated according to \textit{intermediary topics}. 
Intermediary topics are defined as non-conflictive shared interests between users, \ie, interests where two persons of opposing views on sensitive issues could communicate and discuss without facing challenging information in a first encounter. 
According to the primacy effect in impression formation \cite{asch1946forming}, first impressions matter, making such intermediary topics important when introducing people.
In recommender systems, recommendations based on intermediary topics would indirectly address the problem of exposing people to others of opposing views in a non-challenging context.

In this work, we extend the definition of intermediary topics \cite{graells2014people}. In addition, we formally evaluate this redefinition by considering the following research question: \textit{are intermediary topics more diverse in terms of political stances and language than non-intermediary topics?}
We approach this question by performing a case study on the micro-blogging platform Twitter, with users who discussed sensitive issues, \ie, ideological or political themes that would make people reject connecting or interacting with others.
In particular we focus on the analysis of discussion around \textit{abortion} in Chile.
Chile has one of the strictest abortion laws in the world \cite{shepard2007abortion}, yet at the same time a majority of population is in favor of its legalization \cite{CEPsurvey2013}, making it a controversial topic suitable for analysis.
Our contributions include a quantification of the homophilic structure of discussion around this topic in Chile, and a confirmation of the diversity of people \wrt political stances in intermediary topics. 

This paper is organized as follows: after reviewing the background work ($\S$ \ref{sec:background}), we define the methods and concepts needed to study intermediary topics ($\S$ \ref{sec:methods}). Then, we perform a case study in Chile ($\S$ \ref{sec:case_study}). Finally, we discuss results and implications ($\S$ \ref{sec:discussion}).

\section{Background}
\label{sec:background}
\textit{Homophily} is the tendency to form ties with similar others, where similarity is bound to many factors, from sociodemographic to behavioral and intra-personal ones (see a literature review by \citet{mcpherson2001birds}).
In micro-blogging platforms, homophily has been observed in terms of political leaning \cite{barbera2015birds}.
Because of homophily, ego-network structures can help to 
recommend people to interact with \cite{chen2009make,hannon2010recommending}.

In our work, we propose \textit{intermediary topics} as a feature to consider when recommending users to follow. 
The intuition behind intermediary topics is that they focus on homophily in specific shared interests that are non confronting nor challenging, \ie, unlikely to provoke cognitive dissonance.
Our definition of intermediary topics is based on topic modeling using \textit{Latent Dirichlet Allocation} \cite{blei2003latent,ramage2010characterizing}.
In particular, we build a \textit{topic graph} of relations between latent topics, and find which ones are more likely to include people from diverse political backgrounds by estimating the information centrality \cite{brandes2005centrality} of latent topics.

Although topic modeling has been used before to measure homophily by considering user similarity \cite{weng2010twitterrank}, we measure its presence as the deviation from the expected interaction behavior given the population distribution in terms of user stances on specific controversial political issues.  This distinction is important given that homophily also appears in other dimensions (\eg, demography).

To study political leaning in social media, in particular in micro-blogging platforms, the first challenge is to actually detect what is the political leaning of users, as this attribute is not usually part of a public profile.
One way to address the issue of classifying users is through supervised machine learning \cite{conover2011predicting} and bayesian estimation \cite{boutet2013s}, among other methods.
Features used in classification include vocabulary, hashtags, and connectivity with accounts with known political leaning.
Knowing political alignment of users allows to study group polarization.
In a work related to our case study, \citet{yardi2010dynamic} studied debates about abortion in Twitter, in particular between users of \emph{pro-life} and \emph{pro-choice} stances.
Their results indicate that the interaction between users having the same stance reinforced group identity, and discussions with members of the opposite group were found to be not meaningful, partly because the interface did not help in that aspect.
In our work, we focus on a previously unexplored context: a politically centralized Latin-American country \cite{graellsbalancing}.
We complement previous work and help to understand the differences in political discussion around the globe.

\section{Methods}
\label{sec:methods}
In this section we present our methodology to model users' intermediary topics, which extends previous work \cite{graells2014people}.

\spara{Sensitive Issues and Shared Interests}
\textit{Sensitive issues} are political or ideological topics for which their stances or opinions tend to divide people. 
This considers topics like \textit{global warming}, \textit{social security}, \textit{health care reforms}, and \textit{abortion}.
Such topics tend to polarize people, \ie, users who support one stance in abortion do not interact with users who support another stance, a behavior explained by homophily and cognitive dissonance.
Conversely, \textit{shared interests} are topics for which their stances or opinions do not, in normal conditions, tend to divide people.
As example, people who support the soccer team \textit{F.C. Barcelona} have a rivalry with people who support \textit{Real Madrid F.C.}, however, the selective exposure mechanism would not be activated when discriminating information coming from people who support the opposite team--in fact, in some cases, they might be interested in such information.
Other contexts can be less challenging as there might be no explicit rivalries.
For instance, people with different musical tastes might be interested in discussing the particularities of their preferred music styles for comparison with others.
As such, those shared interests could be good features to consider when introducing people \cite{graells2014people}, specially when considering first impressions \cite{asch1946forming}.

\spara{Representation of User Stances in Sensitive Issues}
An assumption we make \wrt user stances is that they are linked by partisan political ideology, \eg, conservative/liberal people share views on different sensitive issues.
Then, to estimate user stances, we first need to be able to estimate what users say \wrt sensitive issues. 
In Twitter, often users annotate their tweets with \emph{hashtags}, which are text identifiers that start with the character \texttt{\#}. 
For instance, \textit{\#prochoice} and \textit{\#prolife} are two hashtags related to two abortion stances, and each one of those stances has specific words related to them (\eg, \textit{``right to choose''} is pro-choice, and \textit{``it is life since conception''} is pro-life). 
\citet{pennacchiotti2011democrats} call those related words \textit{prototypical words and hashtags}. We refer to both as prototypical keywords indistinctively.
For any sensitive issue under consideration, we collect relevant tweets based on prototypical keywords (\eg, \textit{\#prochoice}, \textit{\#prolife}, \textit{abortion}, \textit{pregnancy}, \textit{interruption}, etc.). 
Those keywords can be extracted from a knowledge base of issues, with their respective related stances and associated terms.
This knowledge base should be manually constructed to account for the social context of the population under study, as well as the contingency surrounding political discussion.

We build \textit{user documents}, defined as the concatenation of tweets from each user.
We represent each user document $u$ as a vector
\[\vec{u} = [w_{0}, w_{1}, \ldots, w_{n}] , \]
where $w_{i}$ represents the vocabulary word $i$ weighted using TF-IDF~\cite{baeza2011modern}:
\[w_{i} = freq(w_{i}, u) \times \log_{2} {|U| \over |u \in U: w_{i} \in u|} ,\]
where $U$ is the set of users, and the vocabulary contains all prototypical keywords as well as all other words used by them.
Note that the user document can be built with all tweets and retweets for each user, as well as a subset of both.
In particular, we consider tweets and retweets, but not replies to other users, as they are less likely to contribute information to the document. 
Likewise, for each issue stance we build a stance vector $\vec{s}$, defined as the vectorized representation of tweets containing its prototypical keywords:
\[\vec{s} = [w_{0}, w_{1}, \ldots, w_{n}] ,\]
with $w_{i}$ weighted according to TF-IDF \wrt the corpus of user documents.

Using these definitions we can estimate how similar is the language employed by a specific user with the known stances of a specific issue.
Formally, we define a user stance \wrt a given sensitive issue as the feature vector $\vec{u}_{s}$ containing the similarity of user $\vec{u}$ with each issue stance.
In this way, we consolidate all similarities in a \textit{user stance vector}:
\[\vec{u}_{s} = [f_{0}, f_{1}, \ldots, f_{|S|}] ,\]
where $S$ is the set of stances for the all sensitive issues under consideration, and $f_{i}$ is the cosine similarity between $\vec{u}$ and the issue stance $\vec{s}_{i}$:
\[
\text{cosine\_similarity}(\vec{u}, \vec{s}_{i}) = \frac{\vec{u} \cdot \vec{s}_{i}}{\parallel \vec{u} \parallel  \parallel \vec{s}_{i} \parallel }
. \]

Having this representation of user stances, we define the \textit{view gap} \wrt a sensitive issue between two users as the distance between their respective user stance vectors.

\spara{Topic Graph}
To build the topic graph, we rely on \textit{Latent Dirichlet Allocation}. 
LDA is a generative topic model that clusters words based on their co-occurrences in documents, and defines latent topics that contribute words to documents.
In the past, this model has given reliable results when applied to user documents.
Thus, by using LDA we are able to estimate $P(t \mid u)$, for a given latent topic $t$ and a given user document $u$ from the set of users $U$.
The topic graph is an undirected graph $G = \{T, V\}$, where the node set $T = \{t_0, t_1, \ldots, t_k\}$ is comprised of the $k$ latent topics obtained from the application of LDA to the user documents, in the same way as \citet{ramage2010characterizing}.
The edge set is defined as $V = \{v_{i,j} \colon P(t_i \mid u) \geq \epsilon \wedge \; P(t_j \mid u) \geq \epsilon \; \exists \; u \in U\}$, \ie, two nodes are connected if both corresponding topics contribute (with a minimum probability of $\epsilon$) to the same user document.
Note that edges are weighted according to the fraction of user documents that contributed to it.

\spara{Intermediary Topics}
To estimate which topics are suitable to be used for recommendation of people of opposing views, we estimate the centrality of each node in the topic graph.
In contrast to a previous definition of intermediary topics \cite{graells2014people}, instead of \textit{betweenness centrality} we compute \emph{current flow closeness centrality} \cite{brandes2005centrality} of nodes, which is equivalent to \textit{information centrality} \cite{stephenson1989rethinking}. 
If the topic graph is considered as an electrical network, with edges replaced with resistances, information centrality is equivalent to the inverse of the average of correlation distances between all possible paths between two nodes.
Using this analogy, we expect to measure the degree in which a topic might represent a shared non-challenging interest (\ie, those with the least resistance) between two users. 
Hence, we redefine \emph{intermediary topics} as topics whose centrality is higher than the median centrality of the entire graph.

In the next section, we evaluate this methodology through a case study of political discussion in Chile. 

\section{Case Study: Abortion in Chile}
\label{sec:case_study}
In this section we describe a case study where we analyze the issue of abortion in Chile using our methodology.
In the context of on-going campaigns for presidential elections, we crawled tweets from July 24th, 2013 to August 29th, 2013 using the \emph{Twitter Streaming API}.
Although we crawled tweets about general political discussion, we did focus crawling and analysis on \textit{abortion}.
After the analysis, we statistically evaluate intermediary topics to find how they differ in comparison to non-intermediary topics. 

\spara{Why Abortion in Chile? The Duality in Discussion}
The history of abortion in Chile is long, being declared legal in 1931 and illegal again in 1989. 
As of 2015, abortion is still illegal, making Chile one of the countries with most severe abortion laws in the world \cite{shepard2007abortion}. 
Abortion in Chile as a sensitive issue has good properties for analysis, as it is constantly being discussed in the political active population.
On one hand, 61\% of population was estimated to be catholic, and 21\% professed another religion, while only 19\% of the population was atheist or agnostic \cite{adimarksurvey}.
On the other hand, 63\% of the Chilean population was in favor of legalization of abortion in 2013 \cite{CEPsurvey2013}.
The occurrence of several protests around public education, same-sex marriage and abortion, among other sensitive issues, are encouraging the usage of micro-blogging platforms and social networks to spread ideas and generate debates (for a discussion on the student movement in Chile see \citet{barahona2012tracking}).
There is a duality in how the country approach political issues. On one hand, a majority of the population is estimated to have conservative views.
On the other hand, a majority of population is in favor of legalization of abortion.
Because a growing portion of the population is asking for reforms using social media as a primary communication and organization device, Chile is an ideal scenario for our analysis.

\subsection{Dataset Description}

\begin{figure}[tb]
\centering
\includegraphics[width=\linewidth]{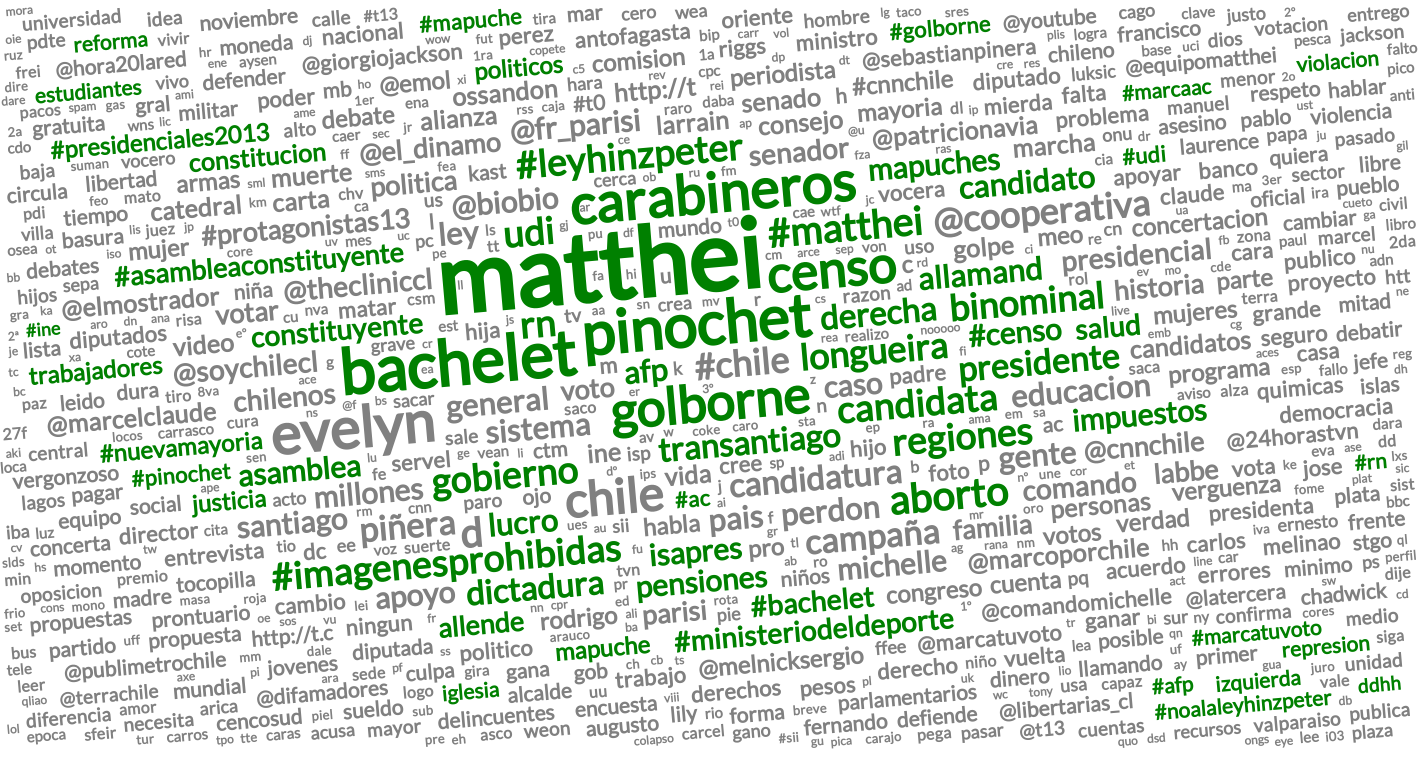}
\caption{Wordcloud of frequent terms in the collection. Green terms were used as query keywords for crawling. Font size is proportional to frequency.}
\label{fig:wordcloud_pilot_study_terms}
\end{figure}

\spara{Query Keywords and Filtering}
Initially, we used \emph{query keywords} about known sensitive issues and hashtags: \emph{abortion} (issue), \emph{education} (issue), \emph{same-sex marriage} (issue), \textit{Sebastián Piñera} (president in 2013), \emph{Michelle Bachelet} (candidate), \emph{Evelyn Matthei} (candidate), among others.
When crawling tweets we considered keywords about general political discussion and other sensitive issues (in addition to abortion) because we will consider the relationship between language usage and user stances.
We also added emergent hashtags related to news events that happened during the crawling period. For instance, \textit{\#yoabortoel25} is about a protest held on July 25th \cite{canelo2015control}. 
Figure \ref{fig:wordcloud_pilot_study_terms} shows the most frequent terms found in our collection.
The most prominent words are last names of candidates, namely \textit{Evelyn Matthei}, \textit{Michelle Bachelet}, \textit{Pablo Longueira} and \textit{Laurence Golborne}. 
The last name of the dictator \textit{Augusto Pinochet} is also prominent.
Other prominent keywords are \textit{carabineros} (the police), \textit{censo} (the national level census conducted in 2012, with multiple flaws discovered in 2013), \textit{Transantiago} (public transport system in Santiago), \textit{isapres} (the private health system) and \textit{AFP} (the name of the Chilean private pension system, composed of several \textit{Administrators of Public Founds}).
We filtered tweets in other languages than Spanish, tweets that were not geolocated to Chile according to users' self-reported location, as well as tweets about unrelated themes.

\spara{Dataset Size}
In total, we analyzed 367,512 tweets about political discussion from 57,566 accounts that were geolocated in Chile using a gazetteer.
Of those tweets, 18,148 are related to abortion, as they contain at least one prototypical keyword (see Table \ref{table:abortion_documents} for the list of keywords related to abortion).
The vocabulary size is 38,827, filtering out all keywords that appear in less than 5 tweets.

\begin{\bigtable}[tb]
\centering
\footnotesize
\caption{Keywords used to characterize the pro-choice and pro-life stances on abortion. General keywords plus stance keywords were used to find people who talked about abortion in Twitter. Seeds are users who published tweets with keywords from only one abortion stance.}
\begin{tabulary}{\textwidth}{l|rr|R}
\toprule
Stance & Tweets & Seeds & Keywords  \\
\midrule
\emph{Pro-choice} & 95,173 & 1,934 & {\#abortolibre}, {\#yoabortoel25}, {\#abortolegal}, {\#yoaborto}, {\#abortoterapeutico}, {\#proaborto}, {\#abortolibresegurogratuito}, {\#despenalizaciondelaborto}, \#abortoetico, \#abortolegal, \#abortosinapellido, \#derechoadecidir \\
\emph{Pro-life} & 10,040 & 338 & {\#provida}, {\#profamilia}, {\#abortoesviolencia}, {\#noalaborto}, {\#prolife}, {\#sialavida}, {\#dejalolatir}, {\#siempreporlavida}, {provida}, {\#nuncaaceptaremoselaborto}, \#chilenoquiereabortos, \#conabortonohayvoto, \#yoasesinoel25, \#somosprovida \\
General Words & -- & -- &  {aborto(s)}, {abortista(s)}, {abortados(as)}, {abortivo(a)}... \ldots (tenses of \textit{to abort} in spanish) \\
Related Hashtags & -- & -- &  {\#marchaabortolegal}, {\#bonoaborto}, \#cifrasaborto, \#feminismo \\
Relevant Accounts & -- & -- &  @elardkoch, @siemprexlavida, @quieronacer, @mileschile, @melisainstitute, @ObservatorioGE \\
Contingency Words & -- & -- &  terapéutico, violada, violación, violaciones, interrupción, inviabilidad, embarazo, embarazada, feto, embrión, fecundación, antiaborto, feminismo \\
\bottomrule
\end{tabulary}
\label{table:abortion_documents}
\end{\bigtable}

\spara{Pro-Choice and Pro-Life Stances}
We manually built a list of words, accounts, and hashtags related to abortion and its two stances. 
We iteratively explored the dataset to find co-occurrences of prototypical keywords like \textit{abortion}, \textit{\#abortolibre} (\textit{free abortion}) and \textit{\#noalaborto} (\textit{no to abortion}).
For pro-choice and pro-life keywords, the number of seed users and their number of tweets are displayed. 
These seeds represent whether a user document contained keywords from one stance but not from the other, \eg, a user document that contains at least one pro-choice keyword and no pro-life keywords is considered a pro-choice seed user.
As observed in Table 1, the number of pro-choice seed users outnumbers those of pro-life stance (1,934 pro-choice against 338 pro-life).
This does not necessarily indicate the proportion of users from both stances. 
For instance, after performing a manual exploration, some pro-life users who identify themselves as pro-life in their biographies, tend to inject content into pro-choice timelines by publishing tweets with prototypical hashtags from the opposite stance \cite{conover2011political}. 

\begin{figure}[tb]
\centering
\begin{subfigure}[b]{0.49\linewidth}
        \includegraphics[width=\linewidth]{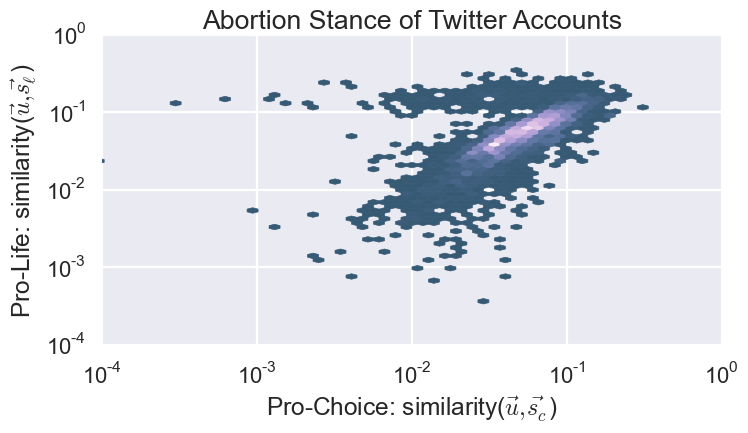}
\end{subfigure}%
\begin{subfigure}[b]{0.49\linewidth}
        \includegraphics[width=\linewidth]{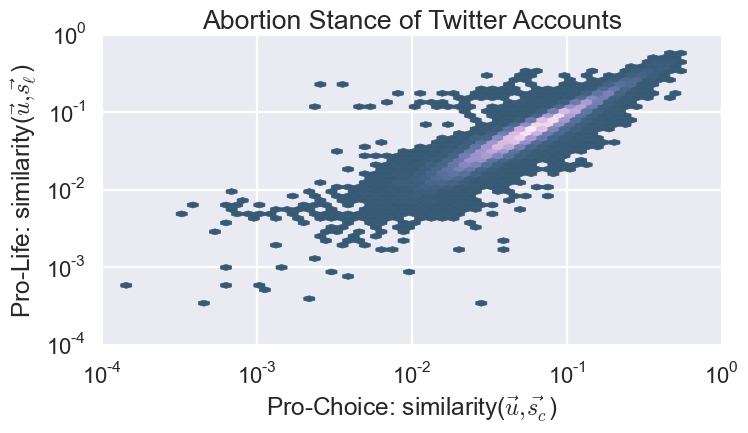}
\end{subfigure}%
\caption{Distributions of user stances based on similarity between user vectors and stance vectors (pro-life and pro-choice). Left: stances of users who tweeted about abortion. Right: stances of all users in the dataset. Both charts use a log-log scale.}
\label{fig:abortion_study_user_stances}
\end{figure}

To build the stance vectors of pro-choice and pro-life stances, we concatenated the tweets of the corresponding seed users of each stance.
Then, according to our methodology, we estimated the user stances on abortion by computing the cosine similarity between user vectors and the stance vectors.
These similarities are displayed with hexagonal binning in Figure \ref{fig:abortion_study_user_stances}, where the $x$ axis represents similarity with the pro-choice stance vector $\vec{s_{c}}$; and the $y$ axis represents similarity with the pro-life stance vector $\vec{s_{\ell}}$. 
We display two charts: one for users who have tweeted about abortion (8,794) on the left, and one that considers all users on the dataset (57,566) on the right.
This is possible because the user stance vectors are constructed using all the vocabulary employed by seed users; hence, they contain valid weights for words unrelated to abortion, but related to additional issues that those users discussed.
Under the assumption that sensitive issues have a degree of correlation among stances in different issues, this allows us to estimate a tendency for all users.
%
We define \textit{stance tendency} as:
\[\text{tendency} = \text{cosine\_similarity}(\vec{u}, \vec{s_{c}}) - \text{cosine\_similarity}(\vec{u}, \vec{s_{\ell}}) . \]
We classify users with $\text{tendency} \ge 0$ as pro-choice, and pro-life otherwise.
The median stance tendency is 0.02, showing a slight tendency towards the pro-choice stance: 54.98\% of users are classified as pro-choice, while 45.02\% of users are classified as pro-life. 
Pro-choice users published 10.24 tweets in average, while pro-life users published 10.48 tweets in average. 

According to the \textit{Center of Public Studies} \cite{CEPsurvey2013}, 63\% of the Chilean population was in favor of legalization of abortion in 2013. 
Our predicted proportion of user stances does not differ from expectations according to a chi-square test ($\chi^2 = 2.76$, p $= 0.10$). 
While the Twitter population is not demographically representative of the population, this result indicates that abortion stances are reflected on the micro-blogging platform Twitter.

\begin{figure}[tb]
\centering
\includegraphics[width=\linewidth]{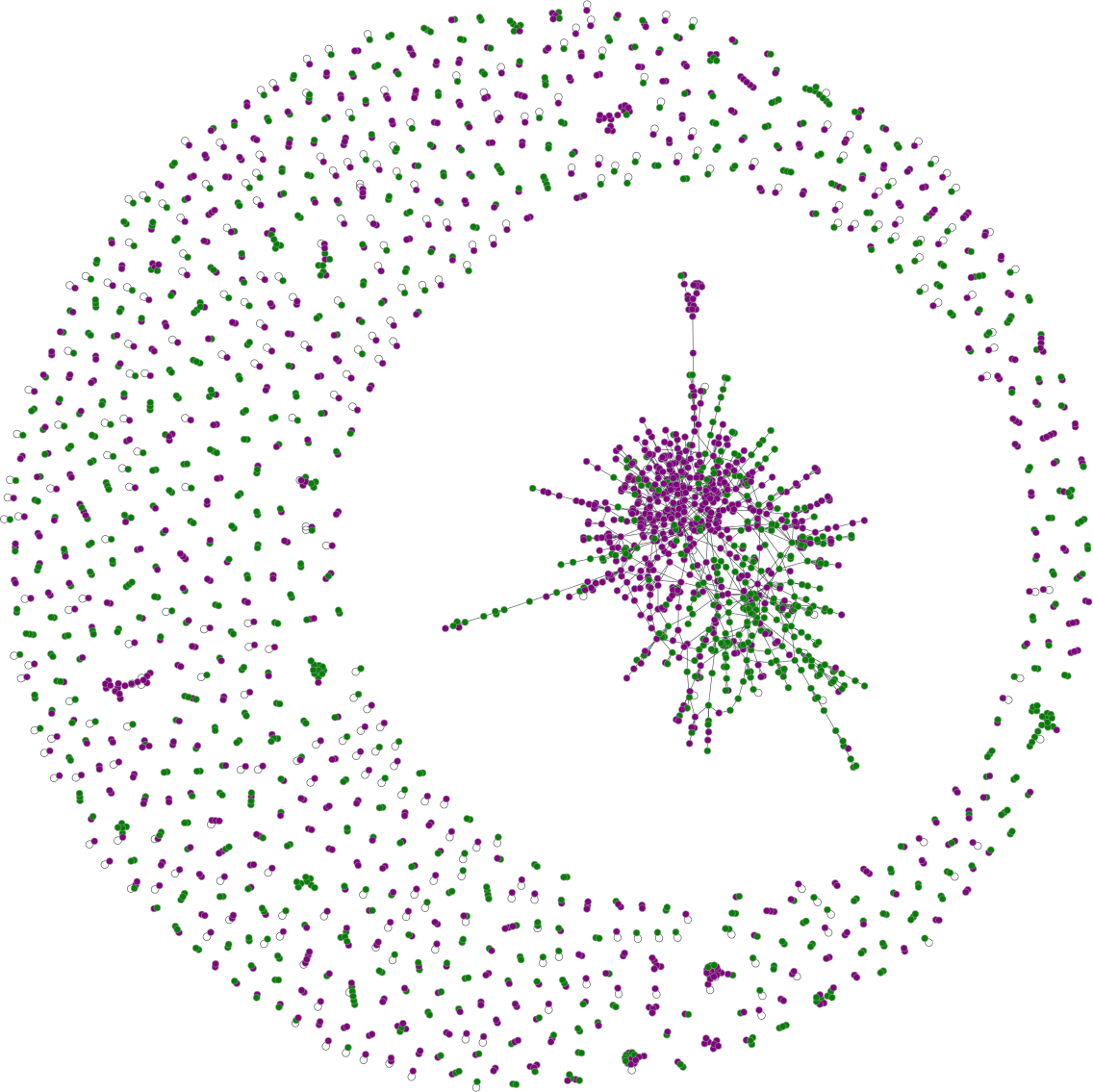}
\caption{A spring-based graph visualization of two-way user interactions in abortion discussion, where nodes are users. Color encodes abortion stance (purple: pro-choice; green: pro-life).}
\label{fig:abortion_two_way_graph}
\end{figure}

\subsection{Homophily in Two-Way Interactions}
Having predicted a stance for each user in the dataset, we are able to evaluate if the interactions in the dataset are homophilic, \ie, we test if users tend to interact with people of the same abortion stance.
To do so, we study 2-way interactions. 
Mentions and retweets are 1-way interactions, where the target user is not necessarily a participant of the interaction.
When the target user replies to the mention or the retweet, we consider it a 2-way, bidirectional interaction.
To measure homophily, we estimate the aggregated interactions between users in both stances, and compare their inter-stance proportions with the proportions of predicted stances for all accounts.
If the interaction behavior is unbiased, then the proportion of interactions between stances should not differ from the proportion of users in each stance.

To avoid bias in the estimation, we only considered each pair ($u_1, u_2$) once per inter-stance interactions.
The number of 2-way interactions found for each stance is: pro-choice, 2,234; pro-life, 2,042.
The structure of those interactions is visualized in Figure \ref{fig:abortion_two_way_graph}, where it can be observed that the largest component has two identifiable clusters, and that small components are prominently of one stance only.
The proportions of interactions with the same stance is similar (pro-choice: 76.45\%; pro-life: 74.24\%).
Given the distribution of user stances, in an unbiased population we would expect that each stance would have bidirectional interactions distributed according to the population, \eg, 54.98\% of pro-choice users' interactions would be with those of the same stance.
A chi-square test indicates that both proportions differ significantly from the expectations (pro-life: $\chi^2 = 29.55$, p $< 0.001$, Cohen's $w = 0.33$; pro-choice: $\chi^2 = 22.91$, p $< 0.001$, Cohen's $w = 0.31$), confirming homophilic behavior in the studied population.

\subsection{Intermediary Topics}
Of all Chileans who published tweets in the case study, we selected a group of 4,077 candidates for analysis of intermediary topics.
We considered users that were likely to be \textit{regular users}, \ie, those who follow less than 2,000 accounts and are followed by less than 2,000 (a limit defined by Twitter).
This filtering was made because regular people are arguably more prone to discuss their own interests, unlike popular accounts which may be from media outlets, blogs, or celebrities.
From those users, we crawled 1,400,582 tweets from December 6th, 2013 until January 3rd, 2014.  
Jointly with our abortion stance estimation of those users, this makes this dataset useful to test the political diversity of intermediary topics.

We ran LDA with $k = 200$ (a value used before in similar contexts \cite{ramage2010characterizing}), built the topic graph and estimated information centrality as defined by our methodology.
After removing junk topics, which do not contribute to any user document, the graph contains 198 nodes and 6,906 edges.
The median centrality is $1.23 \times 10^{-4}$, and its maximum value is $1.64 \times 10^{-4}$.

\begin{figure}[tb]
\centering
\includegraphics[width=\linewidth]{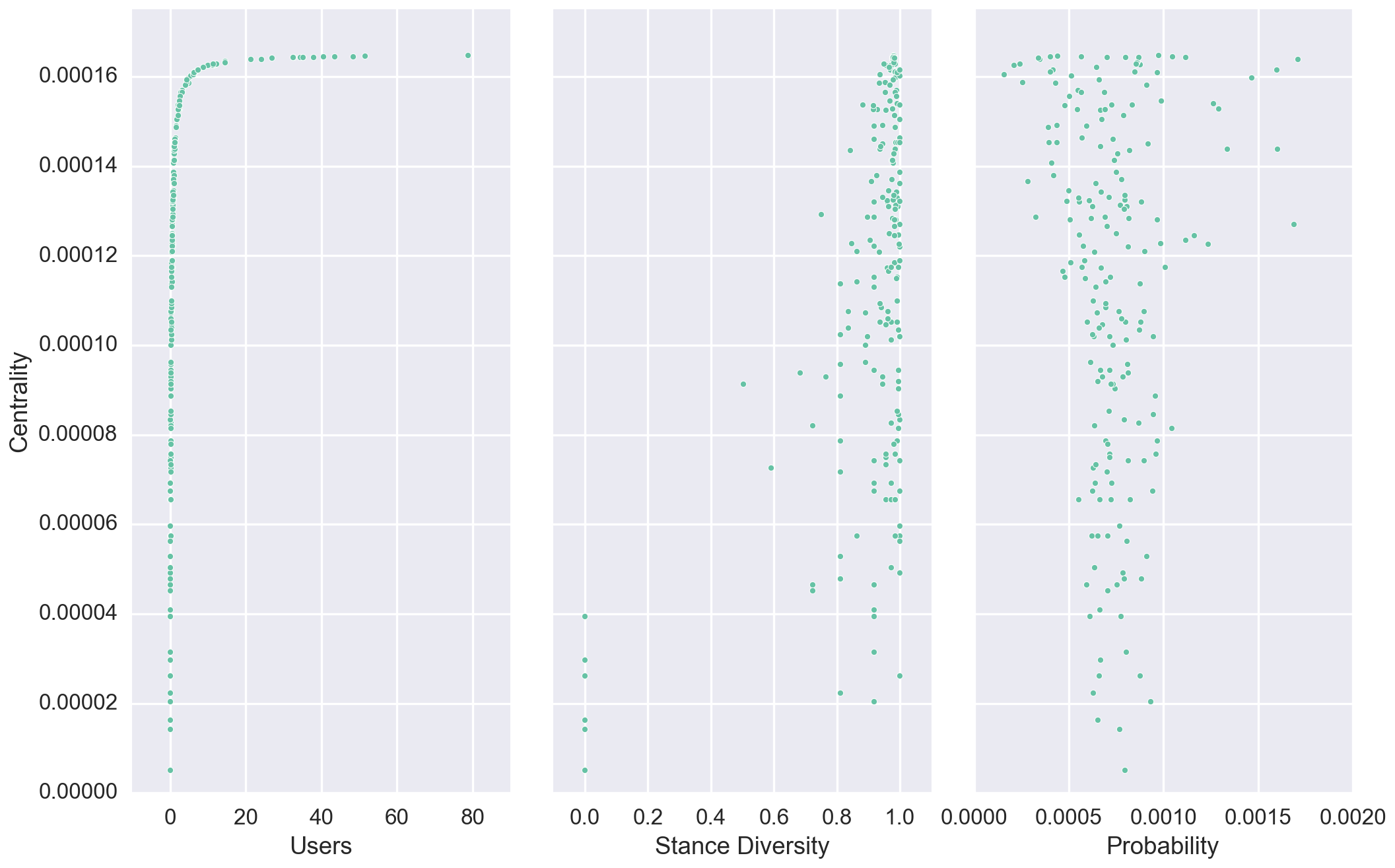}
\caption{Relationship between topic information centrality \cite{brandes2005centrality} and the percent of users the topic contributes to (left), the abortion-stance diversity estimated with \textit{Shannon entropy} \cite{jost2006entropy} (center), and the probability of abortion-related keywords to contribute to each topic (right).}
\label{fig:intermediary_topics_centrality_relations}
\end{figure}

We analyze three variables and their relation with centrality, as well as their differences between intermediary and non-intermediary topics: the percent of users that each topic contributes to (Figure \ref{fig:intermediary_topics_centrality_relations} Left); the probability of abortion keywords to contribute to each topic (Figure \ref{fig:intermediary_topics_centrality_relations} Right), estimated using the LDA model; and the stance diversity (Figure \ref{fig:intermediary_topics_centrality_relations} Center), which is the \emph{Shannon entropy} \cite{jost2006entropy} \wrt the predicted abortion stances for all users related to a topic:
\[\text{diversity} = \frac{-\sum_{i = 1}^{|S|} p_{i} \ \ln {p_{i}}}{\ln |S|} , \]
where $S$ is the set of stances, and $p_i$ is the probability of stance $i$, estimated from the fraction of users assigned to each stance according to our methodology.

\spara{Proportion of Users}
Central topics have much more users than non-central ones: as the number of users increment, centrality does.
This is confirmed by a Spearman $\rho$ rank-correlation of 0.99 (p~$<~0.001$) between proportion of users and centrality.
The maximum proportion of users a topic contributes to is 78.78\%, the median value is 0.56\% and the mean is 4.13\%.
The mean for intermediary topics is 7.99\%, and for non-intermediary topics 0.26\%.
This difference is significant according to a Mann-Whitney U test ($U = 12.10$, p~$<~0.001$).
Hence, intermediary topics are more populated than non-intermediary topics. This is an expected result, because topic graph construction is based on how topics are related to users.

\spara{Stance Diversity}
Nodes with high stance diversity can have low centrality, but they concentrate in the upper middle of the chart. The maximum diversity of a topic is 1, its median value is 0.97 and its mean is 0.91.
The mean for intermediary topics is 0.96, and for non-intermediary topics 0.86.
This difference is significant according to a Mann-Whitney U test ($U = 3.30$, p $< 0.001$), meaning that intermediary topics are more likely to contain a greater diversity of people with different views on abortion than non-intermediary topics.

\begin{figure}[tb]
\centering
\includegraphics[width=\linewidth]{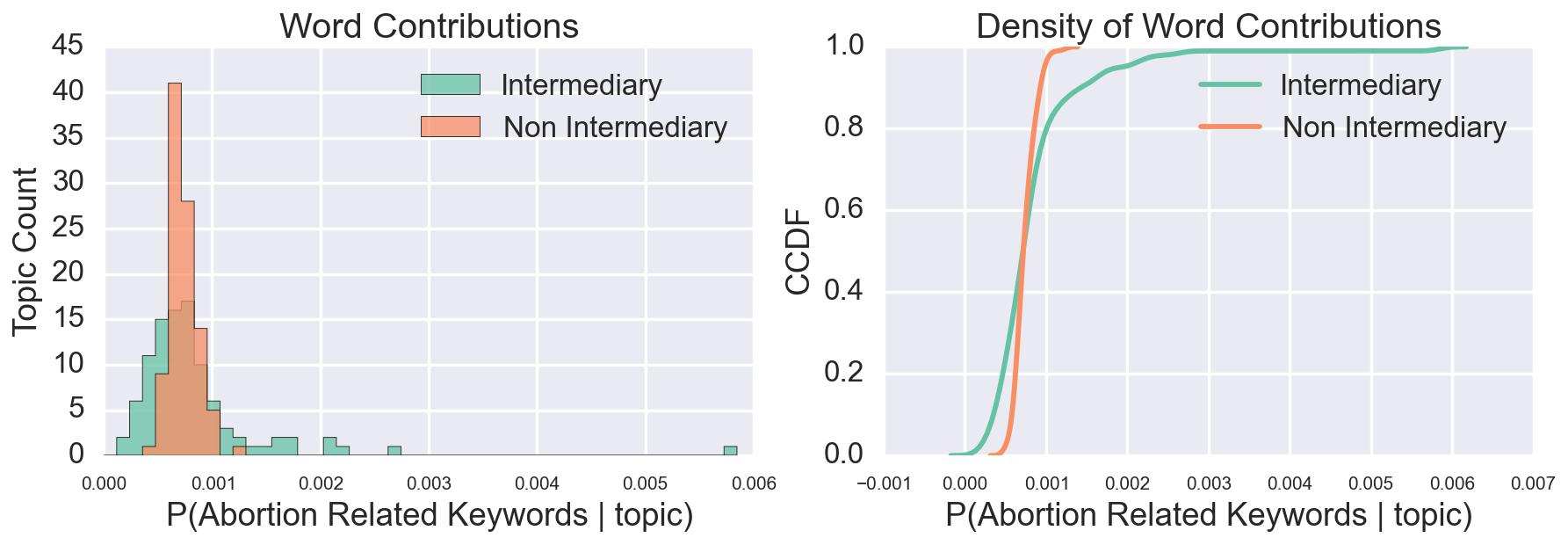}
\caption{Left: Histograms of abortion-related keywords contributions to intermediary and non-intermediary topics. Right: Cumulative Density Function .}
\label{fig:abortion_intermediary_topics_keywords}
\end{figure}

\spara{Topical Probability of Abortion-Related Vocabulary}
Using our set of prototypical keywords, we can estimate the probability of abortion-related vocabulary to contribute to specific topics $P(A \mid t)$, where $A$ is the set of keywords, and $t$ is the target topic:
\[P(A \mid t) = \sum_{i = 1}^{|A|} P(w_i \mid t) , \]
where $w_i$ is the $i$th word in $A$. Note that the LDA model allows us to estimate $P(w_i \mid t)$ directly.
Figure \ref{fig:abortion_intermediary_topics_keywords} displays the distributions and \textit{Complimentary Cumulative Density Functions} (CCDFs) of probabilities for intermediary and non-intermediary topics.
Although the distribution chart hints a potential difference, this difference is not significant according to a Mann-Whitney U test ($U~=~-0.59$, p~$=~0.55$).

\section{Discussion}
\label{sec:discussion}
In this paper we have confirmed that intermediary topics do exist and are measurable. 
We have improved the definition of intermediary topics by \citet{graells2014people}, as well as quantified homophilic discussion and the differences between intermediary and non-intermediary topics. 
In particular, we have found that intermediary topics are more likely to contain a diverse set of users in terms of political stances, and thus, are suitable for use in recommendation of people of opposing views. We devise these topics as important features that could help to avoid \textit{cognitive dissonance} \cite{festinger1962theory} in users when facing recommendations.
Although our results apply to the studied community from Chile, the methods used are generalizable to other communities as long as there are known prototypical keywords for the sensitive issues to be studied.

In addition, the way in which we quantified homophily can be used as a metric to evaluate the polarization in discussion around specific political issues.
In our case study, polarization of stances had considerable effect sizes (measured with Cohen's $w$), meaning that discussion in Chile around abortion is highly polarized, a result supported by national surveys of political discussion \cite{adimarksurvey,CEPsurvey2013}.

A question that arises regarding intermediary topics is: does the definition of intermediary topics hold when considering general political views instead of a specific sensitive issue?
We propose that it does because by definition intermediary topics only rely on the estimation of information centrality \cite{brandes2005centrality}. However, this is left for future work.
Additionally, future work will consider the incorporation of intermediary topics into a recommender system to be evaluated with users, as well as the interaction of intermediary topics with social- and content-based signals.

\spara{Acknowledgments}
We thank the anonymous reviewers for their helpful feedback.
This work was partially funded by Grant TIN\-2012\--38741 (Understanding Social Media: An Integrated Data Mining Approach) of the Ministry of Economy and Competitiveness of Spain.

\printbibliography

\end{document}